\documentclass[twocolumn,showpacs,pra,floatfix]{revtex4}

\usepackage{graphicx}
\usepackage{amsmath}
\usepackage{bm}
\usepackage{psfrag}
\usepackage{color}

\begin{document}

\title{Dipole-dipole interaction between orthogonal dipole moments 
in time-dependent geometries}

\author{Sandra Isabelle \surname{Schmid}}
\email{sandra.schmid@mpi-hd.mpg.de} 

\author{J\"org \surname{Evers}}
\email{joerg.evers@mpi-hd.mpg.de} 

\affiliation{Max-Planck-Institut f\"ur Kernphysik, Saupfercheckweg 1, D-69117
Heidelberg, Germany} 

\date{\today}

\begin{abstract}
In two nearby atoms, the dipole-dipole interaction can couple transitions
with orthogonal dipole moments. This orthogonal coupling accounts for
a number of interesting effects, but strongly depends on the geometry of
the setup. Here, we discuss several setups of interest where the geometry
is not fixed, such as particles in a trap or gases, by averaging
over different sets of geometries. Two averaging methods are compared.
In the first method, it is assumed that the internal electronic 
evolution is much faster than the change of geometry, whereas in the second,
it is vice versa.
We find that the orthogonal coupling typically survives
even extensive averaging over different geometries, albeit with 
qualitatively different results for the two averaging methods. 
Typically, one- and two-dimensional averaging ranges modelling, e.g., 
low-dimensional gases, turn out to be the most  promising model systems.
\end{abstract}



\maketitle

\allowdisplaybreaks

\section{Introduction}

Two nearby atoms can interact in an energy-transfer process via the vacuum  
where one of the atoms is de-excited whereas the other atom is
 excited~\cite{book-agarwal,thiranumachandran,book-ficek, FiTa2002}.
This dipole-dipole interaction has been studied in great detail, albeit
mostly for the case of two-level atoms with parallel transition 
dipole 
moments~\cite{MaKe2003,chang,RuFiDa1995,Ja1993,jump,Fi1991,VaAg1992,entanglement,bargatin,AgPa2001,pra-geometry,breakdown,dfs,strong}. It is known to modify the 
collective system dynamics and thus virtually all observables 
considerably, as was also shown in a number of related 
experiments~\cite{experiment,experiment2,experiment3,hettich,exp-qdot,noel,experiment4}.
Recently, it was found that a new class of effects arises from the
dipole-dipole coupling between transitions with orthogonal dipole 
moments~\cite{AgPa2001,pra-geometry,breakdown,dfs}.
This coupling is somewhat surprising since for single-atom systems, 
only near-degenerate 
non-orthogonal transitions can be coupled via the vacuum~\cite{book-ficek}.
But in real atoms, e.g., transitions 
from one state to different Zeeman-sublevels of a different electronic 
state typically have orthogonal transition dipole moments. 
Therefore, the vacuum-coupling of such transitions in single atoms
usually does not occur,  which is unfortunate, since the corresponding 
couplings are known to give rise to many fascinating
applications~\cite{book-ficek}.

\begin{figure}[b]
\includegraphics[width=3.7cm]{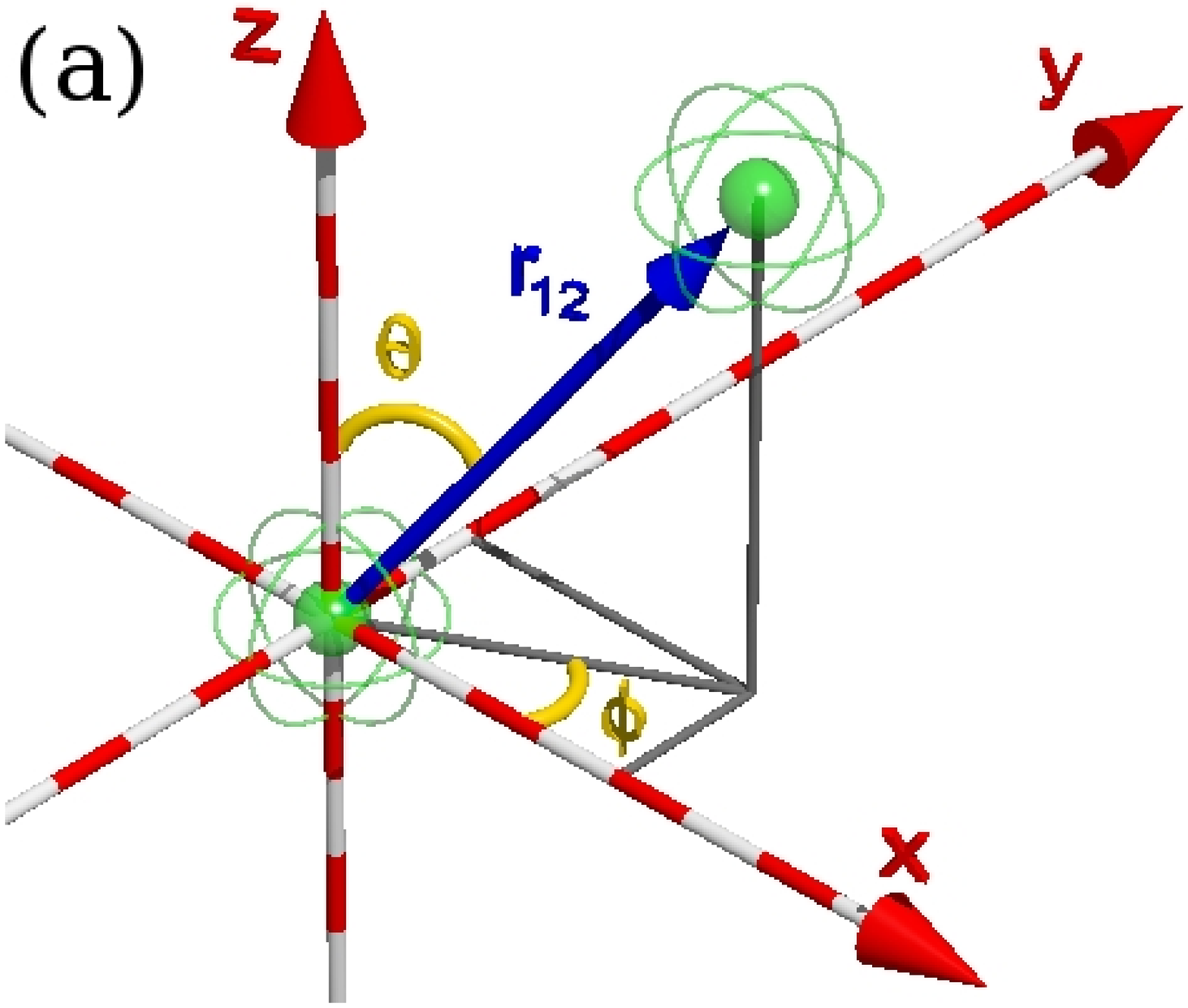}
\hspace*{0.3cm}
\includegraphics[width=4.2cm]{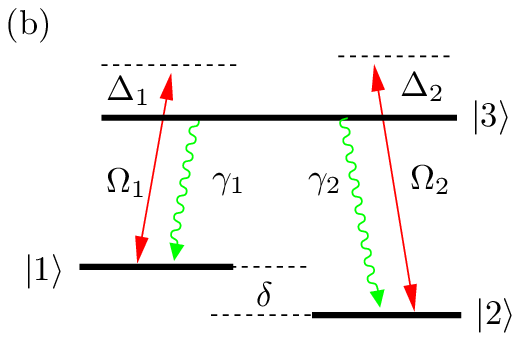}
\caption{(Color online) In the left subfigure the geometry of 
our system is shown. The inter-atomic distance vector $\bm{r}_{12}$ is 
parameterized by the angles $\theta$ and $\phi$ and the length $r_{12}$. 
Atom A is 
located in the origin and atom B at $\bm{r}_{12}$. Each atom is a 
three-level system in $\Lambda$ configuration (b). The two lower states 
have an energy separation $\delta$. $\Omega_1$ ($\Omega_2$) is the 
Rabi frequency of the driving laser field coupling to transition 
$1\leftrightarrow 3$ ($2\leftrightarrow 3$) and the spontaneous 
decay rates are $\gamma_1$ and $\gamma_2$.}
\label{fig-system}
\end{figure}

In contrast, orthogonal transition dipole moments
in different atoms do interact via the vacuum, with coupling coefficients
dependent on the relative alignment of the atoms, see Fig.~\ref{fig-system}. 
It was shown in~\cite{AgPa2001} that this
interaction creates coherences involving excited states that 
are not driven by any laser fields.
This observation can be generalized by studying the
two-particle master equation under rotations of the inter-atomic
distance vector~\cite{breakdown}. It was found that because of
the orthogonal couplings, typically complete Zeeman 
manifolds have to be considered in modelling the
dipole-dipole interaction of two atoms, such that the usual
few-level approximation is no longer possible. 
The orthogonal couplings crucially influence the system 
dynamics. For example, the long-time dynamics
of a two-atom system can strongly depend on the relative orientation of the two
atoms~\cite{pra-geometry}. For a suitable laser and detector 
setup, undampened periodic oscillations in the fluorescence intensity 
are observed for some
relative orientations of the two atoms, whereas the system evolves into
a stationary steady state for other relative orientations. The reason
for this geometry-dependence is the structure of the dipole-dipole constants.
If the coupling of orthogonal transition dipole moments vanishes,
then also the oscillations in the long-time limit vanish.

In many situations of interest, however, the geometry is not fixed.
For example, in a linear trap, the inter-atomic distance usually
can be described classically as a sinusoidal oscillation
around a mean distance. In this case, a dependence of the dynamics
on the orientation of the dipole moments relative to the oscillation 
direction can be expected. A gas of atoms corresponds
to a setup where both the orientation and the distance 
of any given pair changes with time.
Thus the question arises, whether the geometry-dependent effects
of the dipole-dipole interaction of orthogonal transition dipole
moments survive an averaging over different
geometries.

Therefore, here we discuss the fluorescence intensity emitted by
a pair of three-level $\Lambda$-type atoms when averaged over sets
of different geometries of interest, see Fig.~\ref{fig-system}. 
Our primary interest is the
question whether the dipole-dipole couplings of orthogonal
transition dipole moments survives an averaging over different
geometries and thus also is of relevance if the two atoms are
not fixed in space. Since the modulations in the fluorescence
intensity are a direct consequence of these couplings, they are
a convenient indicator and allow for a quantitative analysis.
The second major question involves the way the averaging should 
be treated theoretically. For comparison, we discuss two different
ansatzes. First, one can assume that the internal electronic dynamics
is much faster than the change of the geometrical setup. On the other
hand, we consider the case where the change on geometry is fast
enough such that the atoms essentially see an averaged interaction potential.
The latter approach for example is used in the context of ultracold
quantum gases to derive the $1/r$ long-range potential from the 
dipole-dipole coupling of parallel dipole moments by averaging over 
all possible orientations of the inter-atomic distance 
vectors~\cite{thiranumachandran}. 

We find that in general the orthogonal couplings can survive
an extensive averaging over different geometries as long
as the inter-particle distance remains small. The magnitude
of the effects in the averaged signal, however, strongly depends
on the averaging range, and also on the averaging method. 
Typically, one- or two-dimensional systems can be expected
to show larger effects of the dipole-dipole coupling.
We also show that the two averaging methods considered
can give very different results when averaged over the
same set of geometries. In most situations, however, the
case where the change in geometry is slow as compared to the
internal dynamics is more favorable.

The article is organized as follows: In Sec.~\ref{sec-model},
we present the model system, derive the equations of motion
and discuss our main observable, the time-dependent fluorescence
intensity. In Sec.~\ref{sec-avg}, the two averaging methods
are presented and discussed. Sec.~\ref{sec-result} presents
the results from the averaging for various different situations
of interest. Finally, our findings are discussed and
summarized in Sec.~\ref{sec-summary}.

\section{\label{sec-theory}Theory}
\subsection{\label{sec-model}The model system}
We consider a system consisting of two identical three-level atoms 
in $\Lambda$ configuration, see Fig.~\ref{fig-system}. The atomic 
states have energies $\hbar\omega_i$ ($i\in\{1,2,3\}$).
The transition dipole moments of each individual atom are 
assumed perpendicular, as it is common for near-degenerate electronic 
states in atomic systems such as Zeeman sublevels. For simplicity, both 
transition dipole moments are assumed to be real; the one of the 
$1\!\leftrightarrow\!3$ transition $\bm{d}_1=(d_1,0,0)^T$ is orientated 
along the $x$ direction and that of the $2\!\leftrightarrow\!3$ transition 
$\bm{d}_2=(0,d_2,0)^T$ along the $y$ direction. A comparison with the case 
of complex dipole moments coupling to circularly polarized light 
was given in~\cite{AgPa2001}. It should be noted that it was found 
in~\cite{breakdown} that in general all Zeeman sublevels of two 
nearby dipole-dipole interacting multilevel atoms have to be
considered in order to correctly account for the different 
dipole-dipole couplings occurring in the system. 
Couplings to certain Zeeman sublevels can be eliminated, however, 
in special geometries, or via a detuning between the different 
transition frequencies, thus recovering the well-known few-level 
systems. In the following, we are interested in arbitrary geometries, 
and are thus restricted to an elimination via detunings.
A $\Lambda$-type level scheme could be realized, for example, in a 
four-level $J\!=\!1/2 \leftrightarrow J\!=\!1/2$ scheme~\cite{time-energy} 
subject to a static magnetic field, such that the energy spacing between
the upper states is sufficiently large to neglect dipole-dipole coupling 
to one of the upper states in the four-level scheme.
The frequency difference between the two lower states is denoted 
by $\delta$. Atom A is located in the origin of our coordinate system 
$\bm{r}_{1}=(0,0,0)^T$ and atom B  at $\bm{r}_2=\bm{r}_{12}=r_{12}\,
(\sin\theta\cos\phi, \sin\theta\sin\phi,\cos\theta)^T$, 
where the distance vector between the two atoms is $\bm{r}_{12}$. The 
driving laser fields propagate in $z$ direction. 
For this system the Hamiltonian reads
\begin{equation}
H = H_a + H_f + H_{vac} + H_L\,,
\end{equation}
with
\begin{subequations}
\begin{align}
H_a =&  \sum_{\mu=1}^{2} \sum_{j=1}^{3} \hbar \omega_j \, 
          S_{jj}^{(\mu)} \,, \\
H_f =& \sum_{\bm{k}\lambda}\hbar \omega_{k\lambda}\: 
       a_{\bm{k}\lambda}^\dagger a_{\bm{k}\lambda}\,, \\
H_{vac} =& -\sum_{\mu=1}^{2}\left [ \left ( \bm{d}_1 \,S_{31}^{(\mu)} 
            + \bm{d}_2 \,S_{32}^{(\mu)} \right ) \bm{E}(\bm{r}_\mu) 
	    + \textrm{H.c.} \right ] ,\\
H_L =& -\hbar \sum_{\mu=1}^2 \left ( \Omega_1(\bm{r}_\mu) 
  e^{-i\nu_1 t} S_{31}^{(\mu)} \right. \nonumber \\
  &\left. + \Omega_2(\bm{r}_\mu) e^{-i\nu_2 t} S_{32}^{(\mu)} 
  + \textrm{H.c.} \right )\,.         
\end{align}
\end{subequations} 
$H_a$ represents the free energy of the atomic states. The free energy 
of the vacuum field is described by $H_f$. $H_{vac}$ is the interaction 
Hamiltonian of the vacuum field, and $H_L$ is the term describing the 
interaction with the laser fields in rotating-wave approximation (RWA). 
The laser fields have amplitudes $\mathcal{E}_i$,
frequencies $\nu_i$ and polarization unit vectors $\hat{\bm{\epsilon}}_i$ 
($i\in\{1,2\}$), respectively. $\Omega_i(\bm{r}) = \Omega_i \exp[i\bm{k}_i 
\cdot\bm{r}]$ with $\Omega_i = (\bm{d}_i \cdot \hat{\bm{\epsilon}}_i)
\mathcal{E}_i / \hbar$ are the corresponding Rabi frequencies. $\bm{E}(\bm{r})$ 
represents the quantized vacuum field modes. Furthermore, $\omega_{k\lambda}$ 
is the frequency of a vacuum field mode with creation and annihilation 
operator $a_{\bm{k}\lambda}^\dagger$ and $a_{\bm{k}\lambda}$, respectively. 
The energy of the atomic state $|i\rangle$ is $\hbar \omega_i$. 
We define atomic operators
\begin{equation}
\label{sij}
S_{ij}^{(k)} =  |i\rangle_k {}_k\langle j| \qquad (i,j\in\{1,2,3\} 
\textrm{ and } k\in\{1,2\})\,, 
\end{equation}
where $|i\rangle_k$ denotes the $i$th electronic state of atom $k$.
For $i\!=\!j$, Eq.(\ref{sij}) corresponds to a population, whereas 
for $i\!\neq\!j$ it is a transition operator. 

Choosing a suitable interaction picture it is possible to describe 
the system by the master equation for the atomic density operator $\rho$ 
given by~\cite{pra-geometry}

\begin{align}
&\frac{\partial \rho}{\partial t} = 
 - i\sum_{\mu=1}^{2} \sum_{j=1}^{2} \left [\Delta_j S_{jj}^{(\mu)},
 \rho \right ]
\nonumber \allowdisplaybreaks[2] \\
&+ i \sum_{\mu=1}^{2} \sum_{j=1}^{2} \left [ \left (S_{3j}^{(\mu)} 
\Omega_j(\bm{r}_\mu) + \textrm{H.c.} \right) , \rho \right ] 
\nonumber \allowdisplaybreaks[2] \\
& - \sum_{\mu=1}^{2} \sum_{j=1}^{2} \Bigl [ 
 \gamma_j \left ( 
 S_{33}^{(\mu)}\rho - 2 S_{j3}^{(\mu)}\rho S_{3j}^{(\mu)} 
 + \rho S_{33}^{(\mu)} \right ) 
\nonumber \allowdisplaybreaks[2] \\
& + \Gamma_j^{dd}  \left ( S_{3j}^{(\mu)} S_{j3}^{(\neg \mu)} \rho - 
2 S_{j3}^{(\neg \mu)} \rho S_{3j}^{(\mu)} + \rho S_{3j}^{(\mu)} 
S_{j3}^{(\neg \mu)} \right ) \Bigr ]
\nonumber \allowdisplaybreaks[2] \\
&+ \sum_{j=1}^{2} \left ( i \Omega_j^{dd} \left [S_{3j}^{(1)}S_{j3}^{(2)}, 
\rho \right ]  + \textrm{H.c.} \right ) 
\nonumber \allowdisplaybreaks[2] \\
&- \sum_{\mu=1}^{2} \left [ \Gamma_{vc}^{dd} \left ( S_{32}^{(\mu)}
S_{13}^{(\neg \mu)}\rho - 2 S_{13}^{(\neg \mu)} \rho S_{32}^{(\mu)}  
\right . \right . \nonumber \\
& \left . \left . \qquad \qquad + \rho S_{32}^{(\mu)}S_{13}^{(\neg \mu)}
\right ) e^{i\Delta t} + \textrm{H.c.} \right ]
\nonumber \allowdisplaybreaks[2] \\
&+ \sum_{\mu=1}^{2} \left ( i \Omega_{vc}^{dd} \left [S_{32}^{(\mu)}
S_{13}^{(\neg \mu)}, \rho \right ]e^{i\Delta t}  + \textrm{H.c.} \right ) 
\,. \label{master}
\end{align}  
Here, the RWA and the Born-Markov approximation were used. The first term, 
which contains the detunings $\Delta_i = \nu_i - (\omega_3 - \omega_i)$ of 
the driving laser fields, appears because of the chosen interaction picture. 
The interaction with the laser fields is expressed by the second summand with 
the Rabi frequencies $\Omega_j(\bm{r}_\mu)$. The contribution containing 
$\gamma_j$ represents the individual spontaneous decay of each transition
in the two atoms. In our case the spontaneous decay rate on transition 
$3\!\leftrightarrow\!j$ is denoted as $2\gamma_j$. The term with 
$\Gamma_j^{dd}$ contains the dipole-dipole coupling between a dipole of 
one atom and the corresponding parallel dipole of the other atom. The 
contribution proportional to $\Omega_j^{dd}$ represents the corresponding 
dipole-dipole energy shift. The interaction between a dipole 
moment of one atom and the perpendicular one of the other atom is described by the 
expression containing the cross coupling constants $\Gamma_{vc}^{dd}$ and 
$\Omega_{vc}^{dd}$. The symbol $\neg \mu$ denotes the other atom than $\mu$, 
e.g., for $\mu=2$ one has $\neg \mu = 1$. Note that the interaction picture 
in Eq.~(\ref{master}) is chosen such that the residual explicit time dependence 
$\exp[\pm i\Delta t]$ which cannot be transformed away is attributed to the 
terms that describe dipole-dipole coupling of orthogonal transition 
dipole moments. This choice is motivated by the physical origin of this 
time dependence, which arises from these orthogonal couplings~\cite{pra-geometry}.
The frequency $\Delta$ is determined by $\Delta = \delta + \Delta_2 - 
\Delta_1 = \nu_2 - \nu_1$.

The spontaneous decay rates are given by
\begin{equation}
\gamma_i = \frac{1}{4\pi\epsilon_0} \frac{2|\bm{d}_i|^2\omega_{3i}^3}{3\hbar c^3}\,,
\end{equation}
and the dipole-dipole coupling constants can be calculated from~\cite{AgPa2001}
\begin{subequations}
\label{couplings-gen}
\begin{eqnarray}
\Gamma_i^{dd} &=& \frac{1}{\hbar} [\bm{d}_i \cdot \textrm{Im}
 (\overset{\leftrightarrow}{\chi}) \cdot \bm{d}_i^* ]\,, \\
\Omega_i^{dd} &=& \frac{1}{\hbar} [\bm{d}_i \cdot \textrm{Re}
 (\overset{\leftrightarrow}{\chi}) \cdot \bm{d}_i^* ] \,, \\
\Gamma_{vc}^{dd} &=& \frac{1}{\hbar} [\bm{d}_2 \cdot \textrm{Im}
 (\overset{\leftrightarrow}{\chi}) \cdot \bm{d}_1^* ]\,, \label{gamma_vc} \\
\Omega_{vc}^{dd} &=& \frac{1}{\hbar} [\bm{d}_2 \cdot \textrm{Re}
 (\overset{\leftrightarrow}{\chi}) \cdot \bm{d}_1^* ] \,.  \label{omega_vc}
\end{eqnarray}
\end{subequations}
In evaluating these coupling constants we have approximated $\omega_{31} \approx 
\omega_{32} \approx \omega_0$. Re and Im denote the real and imaginary part of the 
tensor $\overset{\leftrightarrow}{\chi}$ whose components are given by
\begin{align}
\chi_{\mu\nu}&(\bm{r}_1, \bm{r}_2) =\frac{1}{4\pi \epsilon_0}
 \left [ \delta_{\mu\nu} \left (  
\frac{k_0^2}{r_{12}} + \frac{i k_0}{r_{12}^2} - \frac{1}{r_{12}^3}
\right )\right.    - \nonumber \\[0.1cm]
& \left . \frac{(\bm{r}_{12})_\mu(\bm{r}_{12})_\nu }{r_{12}^2} \left( \frac{k_0^2}{r_{12}} + \frac{3ik_0}{r_{12}^2} 
- \frac{3}{r_{12}^3} \right )\right ]\,e^{i k_0  r_{12}}\,.
\label{chi}
\end{align} 
$\delta_{\mu\nu}$ is the Kronecker delta symbol.
For our choice of the atomic system, the coupling constants 
between orthogonal dipole moments evaluate to ($\eta = k_0\,r_{12}$)
\begin{subequations}
\label{couplings}
\begin{align}
\Gamma_{vc}^{dd} &= -\frac{3}{4}\,\sqrt{\gamma_1 \gamma_2}\,
\sin(2\phi)\,\sin^2 \theta  \nonumber \\
&\qquad \times \left [
\frac{\sin \eta}{\eta} + 3\left ( \frac{\cos\eta}{\eta^2} - 
\frac{\sin\eta}{\eta^3}\right ) \right ] \,,\\
\Omega_{vc}^{dd} &= -\frac{3}{4}\,\sqrt{\gamma_1 \gamma_2}\,
\sin(2\phi)\,\sin^2 \theta \nonumber \\
&\qquad \times\left [
\frac{\cos \eta}{\eta} - 3\left ( \frac{\sin\eta}{\eta^2} + 
\frac{\cos\eta}{\eta^3}\right ) \right ]\,.
\end{align}
\end{subequations}

Our main observable is the total time-dependent fluorescence intensity emitted by the two 
atoms. It is assumed to be measured by a detector placed on the $y$-axis at the point 
$\bm{R}=R\hat{\bm{R}}$ with $\hat{\bm{R}}=(0,1,0)^T$. 
This intensity is proportional to the normally ordered one-time correlation function 
\begin{equation}
I = \langle \,:\! \bm{E}^{(-)}(\bm{R},t)\:\bm{E}^{(+)}(\bm{R},t)\!:\,\rangle\,,
\end{equation}
where $\bm{E}^{(\mp)}(\bm{x},t)$ are the positive and negative
frequency parts of the vacuum field $\bm {E}(\bm{x},t) = \bm{E}^{(-)}(\bm{x},t)
+\bm{E}^{(+)}(\bm{x},t)$. For our arrangement of the detector, the atoms 
and the laser fields the fluorescence intensity reduces to~\cite{pra-geometry}
\begin{align}
I_y =  w_1^2
 \sum_{\mu,\nu=1}^{2} \left \langle S_{31}^{(\mu)}S_{13}^{(\nu)}
 \right \rangle \:
e^{ik_1 \hat{\bm{R}}\cdot\bm{r}_{\mu\nu}}\,,
\label{int-y}
\end{align}
where $w_1=(\omega_{31}^2\,d_1)/(4\pi \epsilon_0 c^2 R)$ is a 
pre-factor that we neglect in the following.


\subsection{\label{sec-avg}Averaging over different geometries}

The master equation Eq.~(\ref{master}) contains an explicit time dependence 
which is determined by the two driving laser field frequencies. Thus, in general, it
cannot be expected that the system reaches a stationary steady state. This was 
demonstrated in~\cite{pra-geometry}, where it was shown that for $\Delta \neq 0$
in general it depends on the relative alignment of the two atoms whether the
system reaches a stationary state or not. For some geometries, the long-time
limit is constant, whereas for other geometries a periodic oscillation  in the
fluorescence intensity is predicted. Since the relative positions of nearby atoms
in many experimental situations of relevance are not fixed, the question arises
whether any time dependence survives when averaging over a set of geometries.
The most obvious example for this is a three-dimensional volume of gas, where
arbitrary relative orientations and distances can be observed. But also other
sets of geometries may be considered. For example, in~\cite{noel}, an essentially
one-dimensional ultracold quantum gas was studied. In this case, an external 
static field can be used to vary the relative alignment of dipoles and the
trap axis.

In the following, we discuss two different approaches for calculating the averaged 
total fluorescence intensity, which is our main observable. 

\subsubsection{The adiabatic case method}

In general, we have to average over the angles $\theta, \phi$ as well as over
the distance $r_{12}$. We discretize the respective interval of each geometric 
parameter in $N_i$ equal steps of size $\Delta_i$, respectively, 
where $i\in\{r, \theta, \phi\}$. 
This gives rise to $N_r N_\phi N_\theta$ different geometries.
For each of these geometries, we evaluate the coupling constants and numerically
integrate the master equation Eq.~(\ref{master}). From this, we obtain
the time-dependent fluorescence intensity
$\left[I_y(t) \right]_{n_r, n_{\theta}, n_{\phi}}$ for this particular geometry
$(n_i \in \{1,\dots,N_i\})$. 
Finally, we average over all
time evolutions of the different geometries using the expression
\begin{subequations}
\label{Mitteln}
\begin{align}
\overline{\left(I_y\right)}(t) &=
\frac{1}{\mathcal{Q}}\sum_{n_{r}=1}^{N_r} \sum_{n_\theta=1}^{N_\theta} 
\sum_{n_\phi=1}^{N_\phi} \: \mathcal{V}_{r,\theta,\phi} \: \left[I_y(t) 
\right]_{n_r, n_{\theta}, n_{\phi}} \,,\\
\mathcal{Q} &= \sum_{n_{r}=1}^{N_r} \sum_{n_\theta=1}^{N_\theta} 
\sum_{n_\phi=1}^{N_\phi} \: \mathcal{V}_{r,\theta,\phi} \,, \\
\mathcal{V}_{r,\theta,\phi} &= \Delta V_r \: \Delta V_\theta(n_r) \: 
\Delta V_\phi(n_r,n_\theta) \,.
%
\end{align}
\end{subequations}
$\mathcal{Q}$ is a normalization constant.
We work in a spherical coordinate system and do not only consider uniform motions of the atoms. Thus an 
appropriate volume element $\mathcal{V}_{r,\theta,\phi}$ has to be 
considered. In the discretized form, the contributions
from the different coordinates are given by 
\begin{subequations}
\begin{align}
\Delta V_r &= \Delta_r \,, \\
\Delta V_\theta (n_r) &= r_{n_r} \Delta_\theta \,,\\
\Delta V_\phi (n_r,n_\theta)&=r_{n_r} \sin(\theta_{n_\theta})\Delta_\phi \,.
\end{align}
\end{subequations}
When we average over one or two parameters only we omit the other summation(s) 
and volume element(s).

This method of averaging describes the experimentally observable 
signal as long as the change of the geometric setup is slow compared to 
the internal dynamics of the system. Then, the internal dynamics adapts to 
its long-time evolution on a timescale much faster than the change of the geometry. 

 In the following, we will call this 
way of averaging the adiabatic case (AC) method because of the slow 
change of the geometry.

\subsubsection{The average potential method}

In our second method of averaging, a different physical situation is considered. Here, the 
change of the geometry is considered fast compared to the internal dynamics. Then,
the time evolution of the atomic system according to the master equation Eq.~(\ref{master})
is not governed by coupling constants corresponding to a particular fixed geometry.
Rather, the atom experiences an averaged coupling constant.
Therefore, in this case,
we start by averaging all coupling constants Eqs.~(\ref{couplings-gen}) over the range of 
geometries considered.
This can be done analytically without a discretization of the averaging range, but again
taking into account an appropriate volume element. 
Then the averaged coupling constants are given by
\begin{equation}
\overline{\mathcal{C}}=\int_0^{2\pi}\int_0^\pi\int_0^{2\pi} \: \mathcal{C}  \: dV_{r} \,
dV_\theta \,dV_\phi\, ,
\end{equation} 
with $\mathcal{C}\in\left\{\Gamma_i^{dd},
\Omega_i^{dd},\Gamma_{vc}^{dd}, \Omega_{vc}^{dd}\right\}$.
In order to average, e.g.,  over a sinusoidally oscillating distance we 
parameterize $r_{12}=r_m+r_a \sin\alpha$ by a mean distance $r_m$ and an oscillation 
amplitude $r_a$. In this case $dV_{r}=d\alpha$, 
$dV_\theta=r_{12} d\theta$ and $dV_\phi=r_{12} \sin\theta d\phi$.  
Then, the master equation is solved and
the fluorescence intensity is calculated using these averaged coupling constants. 
Finally, the time-dependent intensity is plugged into Eq.~(\ref{int-y}).
Since the expression for the fluorescence intensity Eq.~(\ref{int-y}) also depends
on the orientation of the inter-atomic distance vector, we also average this
expression over the same set of geometries. 

In the following, this way of averaging will be referred to as averaged potential 
(AP) method.


\section{\label{sec-result}Results}

We now turn to a numerical study of our system as outlined in the previous
section. Different ranges of averaging will be considered, according
to different setups of interest. In all cases, the two ways of averaging 
the fluorescence intensity will be compared. 
We choose as initial condition both atoms to be 
in state $|3\rangle$ unless noted otherwise.
%

\begin{figure}[t]
\includegraphics[width=7cm]{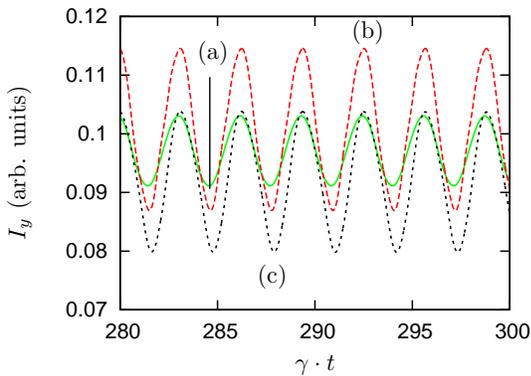}
\caption{(Color online) Time dependence of the fluorescence intensity 
averaged over $r_{12}$ by the AC method. $r_m=0.25\lambda$ and
(a) $r_a=0.02\lambda$, (b) $r_a=0.14\lambda$ and (c) $r_a=0.2\lambda$.
The inter-atomic distance vector is oriented
such that $\phi = \pi/4$ and $\theta = \pi/2$. 
The laser parameters are $\Omega_1 = 3\, \gamma$, 
$\Omega_2 = 5\,\gamma$, $\Delta_1 = 0$, $\Delta_2 = 2\gamma$, and 
the two lower states are assumed degenerate $\delta = 0$.}
\label{r1}
\end{figure}
\begin{figure}[t]
\includegraphics[width=7cm]{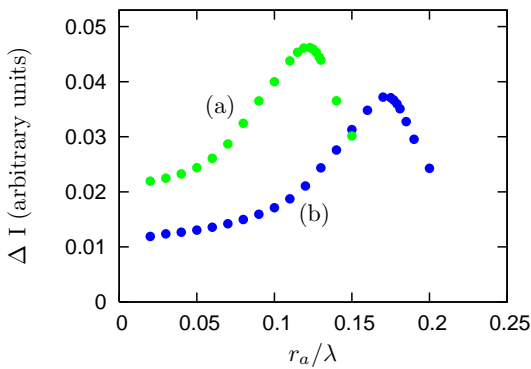}
\caption{(Color online) Dependence of the amplitude of the oscillating 
fluorescence intensity on the oscillation amplitude $r_{12}$ of
the atom for (a) $r_m=0.2\lambda$  and (b) $r_m=0.25\lambda$. 
The other parameters are as in Fig.~\ref{r1}.}
\label{maximaDeltaI}
\end{figure}

\subsection{Averaging over inter-particle distance}
In this section, the orientation of the inter-atomic distance vector is
fixed, while we assume a sinusoidal oscillation of the distance $r_{12}$
around a mean distance $r_m$ with amplitude $r_a$, 
i.e., $r_{12}(\alpha) = r_m + r_a\,\sin(\alpha)$ with $\alpha \in [0,2\pi]$. 
This corresponds to, e.g., atoms in a linear trap. 
In Fig.~\ref{r1}, we choose a mean distance $r_m = 0.25\lambda$ and orientation
$\theta = \pi/2$, $\phi=\pi/4$. The different curves correspond to oscillation
amplitudes $0.02\lambda$, $0.14\lambda$
and $0.2\lambda$, respectively. All curves in this figure were obtained 
using the AC method.
It can be seen that in the averaged signal, the system does not reach a 
steady state in the long-time limit for any of these oscillation amplitudes. 
To analyze the oscillations in the long-time evolution in more detail, we determine the maximum (minimum) 
fluorescence intensity $I_{\textrm{max}}$ ($I_{\textrm{min}}$) in the long-time limit where the intensity
undergoes periodic changes. We define an oscillation amplitude of the 
intensity as $\Delta I=I_{\textrm{max}}-I_{\textrm{min}}$. From Fig.~\ref{r1}, it is clear that $\Delta I$ 
depends on the oscillation amplitude $r_a$. This dependence is depicted
in Fig.~\ref{maximaDeltaI} for small mean distances $r_m$, where it can be seen that 
$\Delta I$ exhibits a resonance in the plot versus the oscillation amplitude $r_a$. 
This resonance can be understood as follows. First, one has to note that as long as the
inter-atomic distance is not too small, typically 
the oscillation amplitude decreases with increasing particle distance, because the 
coupling constants between orthogonal
dipole moments decrease. Therefore, small inter-atomic distances lead to a larger oscillation 
amplitude. Only for very small distances, the oscillation amplitude as well as the total
fluorescence signal are attenuated because the dipole-dipole energy shifts move the atomic transitions
out of resonance with the driving laser field, such that the upper state population is decreased.
This explains why the averaged oscillation amplitude
decreases from the resonance maximum towards smaller oscillation amplitude $r_a$. With smaller 
$r_a$, only larger inter-atomic distances are considered in the averaging, and thus the average 
oscillation amplitude decreases. 
%
\begin{figure}[t]
\includegraphics[width=7cm]{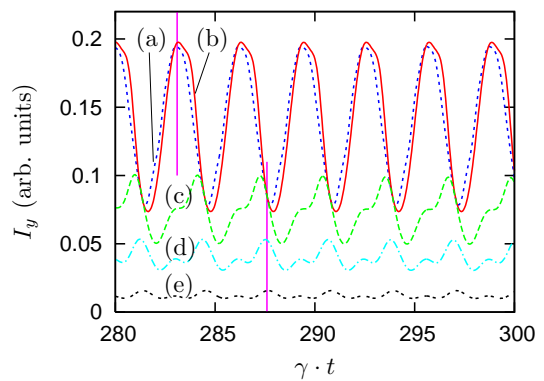}
\caption{(Color online) Time-dependent fluorescence signal for different
fixed distances $r_{12}$ without any averaging. 
(a) $r_{12} = 0.10\,\lambda$, 
(b) $r_{12} = 0.08\,\lambda$, 
(c) $r_{12} = 0.06\,\lambda$, 
(d) $r_{12} = 0.05\,\lambda$, 
(e) $r_{12} = 0.04\,\lambda$. 
The other parameters are as in Fig.~\ref{r1}.
The vertical lines allow to easily
judge the relative phase shifts of the different curves.}
\label{timedep-signal}
\end{figure}
The decrease of the oscillation amplitude $\Delta I$
from the resonance towards higher amplitudes is due to a different mechanism. In Fig.~\ref{maximaDeltaI},
for both mean distances $r_m$, this occurs if $r_a$ is large enough such that inter-atomic 
distances below about $0.06\,\lambda$ are included in the averaging.
Some examples of unaveraged time-dependent signals for different inter-atomic distances are shown
in Fig.~\ref{timedep-signal}. For distances larger than about $0.06\,\lambda$, the relevant
contributions oscillate approximately in phase, see curves (a) and (b) in Fig.~\ref{timedep-signal}.
For smaller distances, however, the contributions move out of phase, as can be seen from
curves (c)-(e). Curves (d) and (e) approximately have maxima where curves (a) and (b) have minima, 
and vice versa. Curve (c) is an intermediate case. Therefore, the oscillations with 
different phases cancel each other in the
averaging process if distances below about $0.06\,\lambda$ are included in the averaging.

\begin{figure}
\includegraphics[width=7cm]{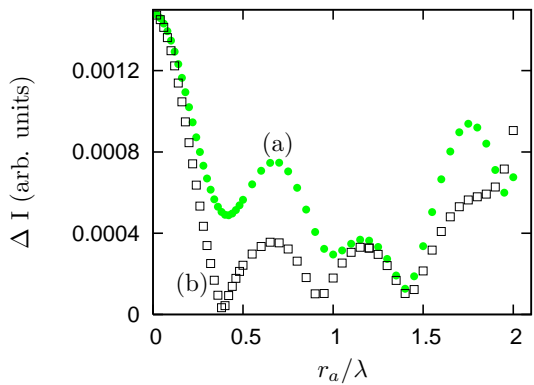}
\caption{(Color online) Dependence of the amplitude of the oscillating 
fluorescence intensity on the oscillation amplitude $r_{12}$ of
the atom for larger mean distance $r_m=2.25\lambda$. 
In (a) we used the AC and in (b) the AP method.
The other parameters are as in Fig.~\ref{r1}.}
\label{DeltaI2}
\end{figure}

In Fig.~\ref{DeltaI2}(a) we show $\Delta I$ in dependence of $r_a$ for a larger 
mean distance $r_m=2.25\lambda$, and over a broader range of oscillation amplitudes. 
It can be seen that the curve exhibits a series of resonances similar to the one
shown in Fig.~\ref{maximaDeltaI}. These again occur due to an alternating 
destructive and constructive superposition of the different oscillations in the averaging
process. The overall amplitude $\Delta I$, however, is small because of the 
overall larger
inter-atomic distances considered in this figure.

%
\begin{figure}[t]
\includegraphics[width=8cm]{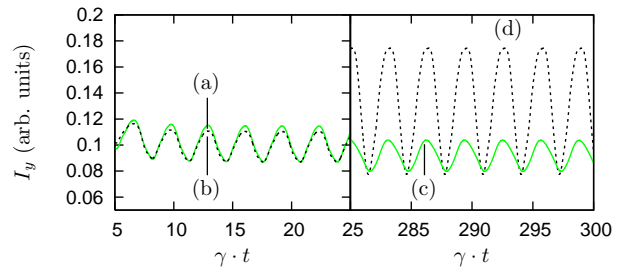}
\caption{(Color online) Comparison of the time dependence of the fluorescence averaged over r by 
both methods. The orientation angles $\theta=0.5\pi$
and $\phi=0.25\pi$ are fix and $r_m=0.25\lambda$. 
(a) AC method, $r_a=0.14\lambda$,
(b) AP method, $r_a=0.14\lambda$,
(c) AC method, $r_a=0.2\lambda$,
(d) AP method, $r_a=0.2\lambda$.
All other parameters are as in Fig.~\ref{r1}.}
\label{rvergleich}
\end{figure}

Finally, we discuss the time-averaged intensity obtained from the
AP method of averaging. Some examples are shown in Fig.~\ref{rvergleich}.
Curves (a) and (c) show our results from the AC method and (b) and (d) those from the AP method. 
The oscillation amplitudes are $r_a=0.14\lambda$ and $0.2\lambda$, respectively. All other parameters
are as in Fig.~\ref{r1}. The left panel shows values for $5\leq \gamma\cdot t\leq 25$ since
the stationary oscillation is reached rapidly for these parameters.
Also in the AP case, the fluorescence
intensity undergoes periodic changes in the long-time limit, see Fig.~\ref{rvergleich}. 
For small oscillation amplitudes $r_a$ there is little difference between the two methods, 
see curves (a) and (b). However, for larger values of $r_a$ the amplitude of the 
oscillations in the AP case is much larger than those obtained in the AC averaging. 
The dependence of the oscillation amplitude on the averaging range for the AP method is 
%
\begin{figure}
\includegraphics[width=7cm]{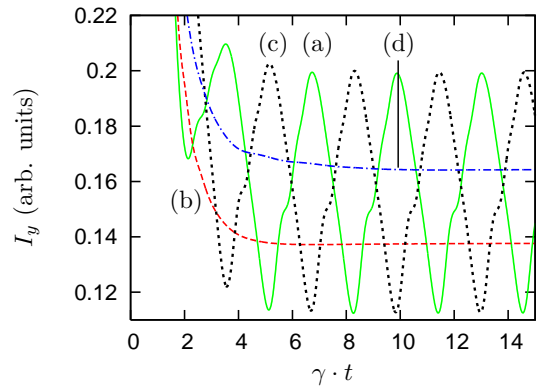}
\caption{Time dependence of the fluorescence averaged over $\theta$ by using the AC 
method. The distance $r_{12}=0.1\lambda$ is fixed. (a) $\phi=0.2\pi$,
(b) $\phi=0.5\pi$, (c) $\phi=0.8\pi$ and (d) $\phi=\pi$.
All other parameters are as in Fig.~\ref{r1}.}
\label{theta1}
\end{figure}
%
shown in 
Fig.~\ref{DeltaI2}, curve (b). As in the corresponding curve (a) for the AC averaging method,
resonance structures appear. But depending on $r_a$, 
the two methods yield either similar or very different oscillation amplitudes.
In addition, the result for the AP method seems to have 
a root at about $r_a = 0.4\lambda$. A careful analysis shows, however, that this minimum is
not a true root. The reason for the minima in the AP curve is that for these oscillation amplitudes,
the turning point at minimum inter-atomic distance is close to a distance where
the coupling constants between orthogonal transition dipole moments are small. 
Then, the averaged coupling constants are small such that the oscillation amplitude has a minimum.
These minima nicely show a crucial difference between the two averaging methods. In the AP
method, it is easy to find averaging ranges where the averaged coupling constants are small or
even vanish. Then, also the oscillation in the long-time dynamics is negligible. 
The results from the AC method,
however, typically remain oscillatory even for such averaging ranges, as the dynamics for each of the
different geometries contributes rather than only an averaged geometry. We will find this 
difference again in the following sections.

\subsection{\label{sec-orientation}Averaging over relative orientation}

In the following, we consider the case where the inter-atomic distance $r_{12}$ is 
fixed, but the relative orientation
and thus the angles $\theta$ and/or $\phi$ are averaged over. 
A realization for this could be a 
Mexican-hat-like potential where one of the atoms is placed in a potential dip in 
the center whereas the other atom is confined to the potential minimum in the rim.
First we fix the angle $\phi$ and assume atom B to move around A on a circle in a plane which 
is perpendicular to the x-y-plane and includes the origin. 
Since $\theta$ is only defined between $0$ and $\pi$ we have to average 
over two semicircles with $\phi$ and $\phi+\pi$ to include the whole circle.
Some examples of our results for $r_{12}=0.1\lambda$ using the AC method of 
averaging are shown in Fig.~\ref{theta1}. Here, the angle $\phi$ is chosen as 
$0.2\pi$, $0.5\pi$, $0.8\pi$ and $\pi$, respectively. 
For $\phi=0.5\pi$ and for $\phi=\pi$ the system reaches a time-independent steady state 
in the long-time limit. That is because the cross-coupling constants are zero for these 
values of $\phi$, as both $\Gamma_{vc}^{dd}$ and $\Omega_{vc}^{dd}$ are proportional to 
$\sin(2\phi)\sin^2(\theta)$, see Eqs.~(\ref{couplings}). 
But even though the coupling constants are zero
in both cases, the resulting intensities are not identical. This demonstrates
that it is not sufficient to analyze the coupling constants alone to understand
the system dynamics.

%
\begin{figure}[t]
\includegraphics[width=8cm]{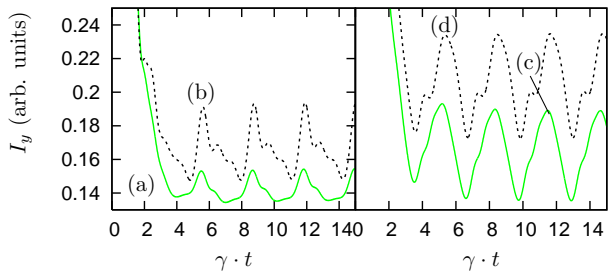}
\caption{(Color online) Time dependence of the fluorescence intensity averaged 
over $\theta$ with fixed inter-atomic distance $r_{12}=0.1\lambda$. 
(a) AC method with $\phi=0.6\pi$,
(b) AP method with $\phi=0.6\pi$,
(c) AC method with $\phi=0.9\pi$,
and
(d) AP method with $\phi=0.9\pi$.
The other parameters are as in Fig.~\ref{r1}.}
\label{thetaVergleich}
\end{figure}
%

In addition we can see from Fig.~\ref{theta1} that there is a phase shift of $\pi$ 
with respect to the oscillation in the long-time limit between 
the two curves for $\phi=0.2\pi$ and $\phi=0.8\pi$. 
We found that in general curves for different values of $\phi$ split
into two groups separated by such a phase shift of $\pi$. 
The first group contains curves for $0<\phi<\pi/2$, whereas the other consists
of curves for $\pi/2 <\phi<\pi$. 
Within each of these groups, the oscillation amplitude of the intensity has the same 
dependence on the angle $\phi$. For $\phi=0$ and $\phi=\pi/2$ the amplitude is zero. 
Then it increases with growing $\phi$ and reaches a maximum for $\phi=0.2\pi$ and
$\phi=0.8\pi$, respectively. Thus, in Fig.~\ref{theta1} the curves with maximum 
oscillation amplitude are shown. 
This separation is likely to appear due to the change of sign of the
cross-coupling 
constants at $\phi=0.5\pi$. This can be understood from the explicit
expressions of the coupling constants Eqs.~(\ref{couplings}).
In the master equation~(\ref{master}) we can see that 
a sign change in the terms with the cross-couplings can be rewritten as 
a constant phase shift factor of $\exp[i\pi]$.  
It is, however, not straightforward to connect this phase shift to
the phase shift seen in Fig.~\ref{theta1}, because the oscillation
frequency of the time-dependent fluorescence in general does not only
depend on $\Delta$, but also, e.g., on the laser field Rabi frequencies.
In addition, one has to note that the geometric parameters $\theta, \phi$ 
enter the total fluorescence 
intensity as well, see Eq.~(\ref{int-y}). 
But our interpretation is further supported by the fact that a change of the inter-atomic distance 
$r_{12}$ has no influence on the separation of our curves into two groups. The separation also
persists for different initial conditions, e.g., atom A in state $|1\rangle$ and 
atom B in $|3\rangle$, and thus is not a consequence of the initial dynamics until the steady state
has been reached.

\begin{figure}[t]
\includegraphics[width=7cm]{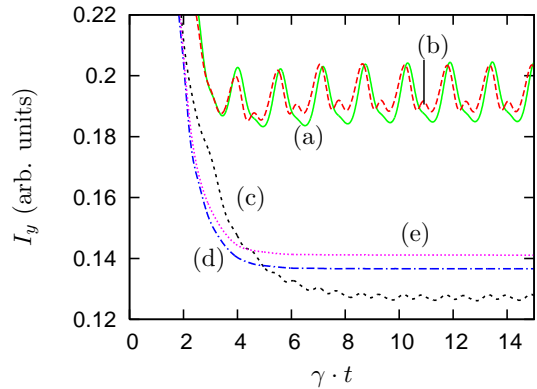}
\caption{(Color online) Time dependence of the fluorescence intensity averaged 
over $\phi$. $r_{12}$ is fixed at $0.1\lambda$. (a) $\theta=0.25\pi$, 
(b) $\theta=0.3\pi$, (c) $\theta=0.5\pi$ and (d) $\theta=\pi$. 
Curves (a-d) are obtained using the AC method. (e) shows a result 
using the AP method for $\theta=0.3\pi$. All other parameters are as in Fig.~\ref{r1}.}
\label{phi}
\end{figure}

Using the AP method the separation into two groups remains, but again the curves 
are different from our results from the other averaging method.
In Fig.~\ref{thetaVergleich} we compare curves from both methods of averaging for 
averaging over $\theta$ with fixed angles $\phi=0.6\pi$ and $\phi=0.9\pi$. 
The curves resulting from the AP 
method have pronounced local extrema in each oscillation period in addition to the global 
ones, and the overall intensity is higher as in the AC case.

Next we assume atom B to move on a circle in the x-y-plane. Thus $\theta$ is fixed 
and we average over the angle $\phi$. The inter-atomic distance is  $0.1\lambda$.
Some examples of our results are shown in Fig.~\ref{phi}, where $\theta$ is chosen 
as $0.25\pi$, $0.3\pi$, $0.5\pi$ and $\pi$.
For the AC method the system does not reach a time-independent state in the long-time 
limit except for the angle $\theta=\pi$.  This is because for this choice of $\theta=\pi$ 
the coupling constants vanish since they are proportional to $\sin(2\phi)\sin^2(\theta)$,
see Eqs.~(\ref{couplings}). 
For any different $\theta$ our system remains oscillating in the long-time limit. 
In case of $\theta=0.3\pi$ one can see local extrema in addition to the global extrema 
in the fluorescence intensity.  In both cases, even the time-averaged average
intensity is considerably larger than in the non-oscillatory case $\theta=\pi$.
Interestingly, for $\theta=\pi/2$, the absolute value of the intensity is lower 
than for the non-oscillatory case $\theta=\pi$. Thus, the orthogonal coupling together
with the averaging can have both an enhancing or a detrimental effect to the total
emitted fluorescence.

For this set of geometries, the AP method of averaging always yields a
stationary long-time limit and thus behaves qualitatively different from
the first method. The reason for this is that the coupling constants 
for the orthogonal couplings vanish upon averaging over then angle $\phi$. 
As discussed before, then the time dependence in the long-time limit also
vanishes, see Fig.~\ref{phi}. 

We now turn to the case of atom B moving around A on a sphere 
with radius $r_{12}$. In this case, neither of the two angles $\theta$ and 
$\phi$ is fixed, and we have to average over both of them while the 
inter-atomic 
\begin{figure}
\includegraphics[width=7cm]{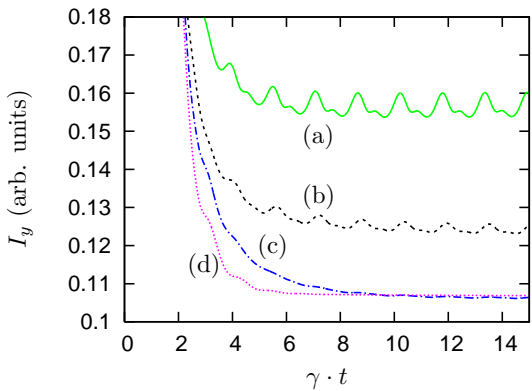}
\caption{(Color online) Time dependence of the total fluorescence intensity averaged 
over $\theta$ and $\phi$ for the inter-atomic distances 
(a) $r_{12}=0.1\lambda$, (b) $r_{12}=0.15\lambda$ 
and (c) $r_{12}=0.2\lambda$. Curves (a-c) are obtained using the AC method.
(d) is the result from the AP method for $r_{12}=0.1\lambda$.
The other parameters are as in Fig.~\ref{r1}.}
\label{phitheta}
\end{figure}
distance is fixed. Some results from both methods are shown 
in Fig.~\ref{phitheta}. We already know that the coupling 
constants vanish when averaged over $\phi$. 
That is why the time dependence in the AP method also vanishes when we 
average over $\theta$ and $\phi$, see curve (d). 
In curves (a)-(c) obtained using the AC method, the  inter-atomic distance is chosen as 
$0.1\lambda$, $0.15\lambda$ and $0.2\lambda$, respectively. One can see that 
both the oscillation amplitude and the absolute value of the fluorescence 
intensity decrease with increasing inter-particle distance.
For the distance $0.2\lambda$ there is almost no oscillation left 
due to the vanishing of the coupling constants with increasing 
inter-atomic distance. This also explains why this curve approaches
the AP method result, where the averaged coupling constants are 
zero. 
As compared to curves (a) and (b) in Fig.~\ref{phi}, 
curve (a) in Fig.~\ref{phitheta} shows that the additional averaging over
$\theta$ does lead to a reduction of the oscillation amplitude.
Still, the oscillation and thus the 
dipole-dipole coupling of orthogonal dipole moments can survive an
averaging over all orientations, depending on the averaging case.


\subsection{Averaging over distance and orientation}

\begin{figure}
\includegraphics[width=7cm]{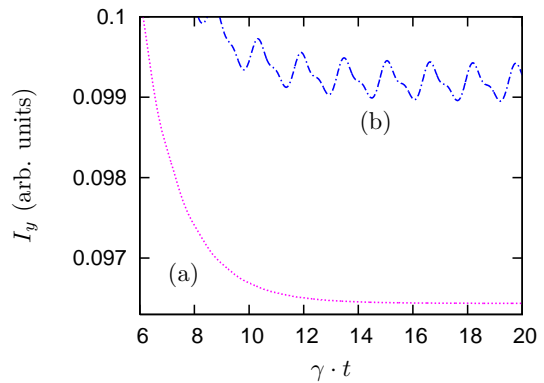}
\caption{(Color online) Time dependence of the total fluorescence intensity 
averaged over r, $\theta$ and $\phi$ for atom B moving on a sphere
around atom A with additional harmonic oscillation of the inter-atomic distance. 
Here, $r_m=0.2\lambda$ and $r_a=0.12\lambda$. In (a) we used the AP and in 
(b) the AC method. All other parameters are as in Fig.~\ref{r1}.}
\label{thetarphi}
\end{figure}

After the individual averaging over the inter-atomic distance and the relative orientation
of the two atoms in the previous sections, we now consider the case of averaging
over both. This situation is realized, e.g., in a gas of atoms, where the relative 
position of any two particles changes with time. An averaging over the two-particle 
configuration space is meaningful, since in a macroscopic volume
of gas at any time there is a finite probability for an arbitrary geometry within
the volume of the sample to be present. A different realization is a sample of
atoms randomly embedded in a host material. In this case,
again an averaging is in order. The two situations differ, however, since the former case
corresponds to a time-dependent geometry for any two-particle subsystem, whereas the
latter case can be represented by a sample of time-independent pairs.

\begin{figure}[t]
\includegraphics[width=6cm]{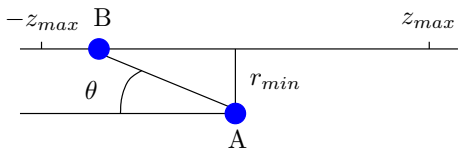}
\caption{(Color online) Geometry for the case of atom B flying past atom A
with constant velocity on a straight line from $-z_{max}$ to $z_{max}$.}
\label{skizze}
\end{figure}

Thus, in the following, we investigate whether in these cases any time dependence of the
fluorescence intensity remains in the long-time limit by considering a system 
where $r_{12}$, $\theta$ and $\phi$ are variable. The three-dimensional case
of course leaves several possibilities for the averaging range. In the following,
we will consider two cases. In the first case, atom B moves on a sphere with 
atom A in its center and additionally oscillates around the mean distance $\bm{r}_{12}$ 
with an amplitude $r_a$. In the second case, the particle fly-by, 
particle B passes atom A moving with constant velocity on a straight 
line, see Fig.~\ref{skizze}.

In Sec.~\ref{sec-orientation} we have seen that averaging the coupling constants 
over $\phi$ makes them vanish, such that the system does not show any time dependence 
in the long-time limit when we use the AP method of averaging. This, of course also
holds true for the three-dimensional averaging for atom B moving on a sphere
with oscillation of the inter-atomic distance. 
In contrast, the AC method of averaging still yields time-dependent fluorescence 
intensities. An example is shown in Fig.~\ref{thetarphi}. Here, the inter-atomic 
mean distance is chosen $r_m=0.2\lambda$ and the oscillation amplitude is 
$r_a=0.12\lambda$. We see that even if we average over all three geometric parameters, 
the system does not reach a time-independent state in the long-time limit, even though
the oscillation amplitude is small.

\begin{figure}[t]
\includegraphics[width=7cm]{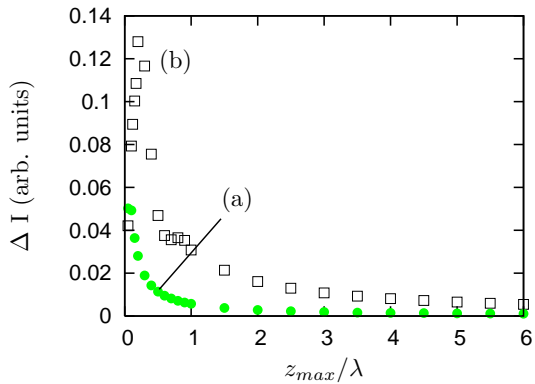}
\caption{(Color online) Dependence of the oscillation amplitude 
of the time-dependent intensity on $z_{max}$ for impact parameter
$r_{min}=0.05\lambda$.  $\phi = \pi/4$, and the other parameters are as in
Fig.~\ref{r1}. In (a) we used the AC and in (b) the AP method.}
\label{x1}
\end{figure}

Finally, we consider the case where atom A flys past atom B along the z-axis 
from $-z_{max}$ to $z_{max}$ with constant velocity, see Fig~\ref{skizze}. 
The angle $\phi$ is fixed and we average over $\theta$ and $r_{12}$ considering 
the respective volume element. We analyzed the case $\phi = \pi/4$ and found
that for both averaging methods the fluorescence intensity remains oscillatory in 
the long-time limit. To further study these oscillations, in Fig.~\ref{x1} 
we show the oscillation amplitude of the time-dependent fluorescence
intensity in the long-time limit against the extend of the motion $z_{max}$.
The  minimum inter-atomic distance is chosen as $r_{min}=0.05\lambda$. 
Curve (a) shows our results from the AP and (b) 
those from the AC method of averaging. 
One can see that in both cases the 
amplitude decreases with increasing $z_{max}$ for large values of $z_{max}$. 
This is because for large distances the dipole-dipole interaction tends to
zero, and oscillations only occur if the particles are close. If
the averaging interval contains increasing ranges of $z$ where there
essentially is no oscillation because of the inter-atomic distance,
then the oscillations in the overall signal decrease.
It is interesting to note, however, that in this averaging configuration
the AP method shown in curve (a) yields much larger oscillations than the 
AC method shown in curve (b). Also, it can be seen that the AP method shows
oscillations over a range of $z_{max}$ up to several wavelengths $\lambda$.
The reason for this is as follows. In the AP method, the coupling constants
are averaged over the different geometries. For $z=0$, the distance between
the particles is $r_{min}=0.05\lambda$. At this position, the coupling
constant $\Omega_{vc}^{dd}$ acquires a large value of more than $330 \gamma$.
Of course, with increasing distance the constant $\Omega_{vc}^{dd}$ rapidly
decreases down to zero. But averaging over a certain range $[-z_{max},z_{max}]$
still gives a considerable averaged coupling constant $\bar{\Omega}_{vc}^{dd}$ 
even for values of $z_{max}$ where the unaveraged coupling constants 
are negligible. This is the reason why the oscillations persist for large
$z_{max}$ values in the AP case. In contrast, in the AC case, contributions
from larger $z$ values do not oscillate at all such that the decrease
of the oscillation amplitude with $z_{max}$ is much more rapid.

\begin{figure}
\includegraphics[width=7cm]{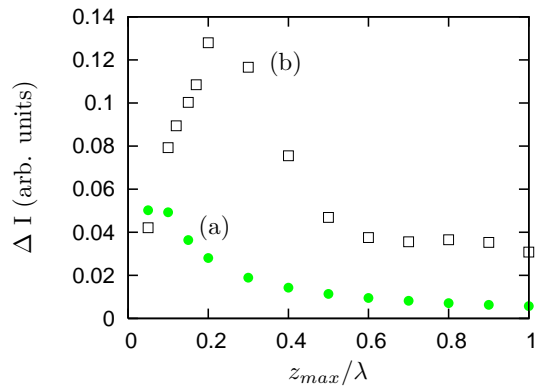}
\caption{(Color online) Dependence of the oscillation amplitude of the 
time-dependent intensity on $z_{max}$ for smaller values of $z_{max}$.
(a) AC method with $r_{min}=0.05\lambda$,
(b) AP method with $r_{min}=0.05\lambda$.
All other parameters are as in Fig.~\ref{x1}. }
\label{x2}
\end{figure}

We now focus on the region with smaller motion extends 
$z_{max}$. The corresponding results are shown in Fig.~\ref{x2}
for $r_{min}=0.05\lambda$.
In the limit $z_{max}\to 0$, the time-dependent fluorescence
approaches the unaveraged curves (d) and (e) in Fig.~\ref{timedep-signal},
which exhibit relatively low oscillation amplitudes. The reason is that
at this small distance, the atomic states are shifted by the dipole-dipole
interaction out of resonance with the laser fields, such that the overall
fluorescence is low. For both methods, the intensity oscillations
first strongly enhance with increasing $z_{max}$, and then decrease
again after passing through a maximum oscillation amplitude.
The AC method results for larger $z_{max}$
essentially remain structureless.
The AP results, however, exhibit some oscillations, 
and only then start to decay monotonously with increasing
averaging range. Due to the complexity of the system, it is difficult
to definitively attribute the oscillation to a property of the system.
We believe, however, that they are due to a similar alternating
constructive and destructive interference in the averaging as the one
that led to the resonance structures in Figs.~\ref{maximaDeltaI} and
\ref{DeltaI2}. Such resonances do not appear in the AC method results,
because there the contributions for higher values of $z$ where the
oscillations in the AP method appear are already too small.

\section{\label{sec-summary}Discussion and summary}
Dipole-dipole interactions between transitions with
orthogonal transition dipole moments gives rise to
a new class of effects in collective quantum systems.
These couplings, however, strongly depend on the geometry
of the setup, and even vanish for some geometries.
Therefore here we have discussed different averaging schemes
to answer the question whether measurable effects
of the dipole-dipole coupling of orthogonal dipole moments
survive if the geometry of the system under study is not fixed.
As observable, we chose the easily accessible
 fluorescence intensity of a pair of
laser-driven $\Lambda$-type atoms, which for suitable 
laser parameters is known to exhibit
periodic oscillations in the long-time limit due to the
orthogonal couplings.

As a main result, we found that the effects of the dipole-dipole
coupling of orthogonal transition dipole moments can survive
extensive averaging over all three spatial dimensions.
We have analyzed the obtained averaged signals,
and expect our physical interpretations to carry over
to other atomic level structures.
Depending on the averaging range, both constructive
and destructive superpositions of the contributions for
the respective geometries is possible, such that a wide
range of results was observed. The results also strongly
depend on the method of averaging, and thus on the
physical situation considered. Typically, the adiabatic
case, where the geometry changes slowly as compared to the
internal dynamics, is more favorable since it better 
preserves the intensity oscillations. In the average
potential case, where the change of geometry is so fast
that the atoms effectively see a dipole-dipole interaction
averaged over the different geometries, some averaging ranges
lead to an exact vanishing of the coupling constants. This usually
does not occur in the adiabatic case.
A somewhat different situation was found in the particle
fly-by, where the averaging over the coupling constants
in the AP method led to a much wider range of distances
over which an effect of the orthogonal couplings can
be observed. 
In general, our results show that the most pronounced 
effects of the orthogonal
couplings in systems with variable geometry can be expected
in one- or two-dimensional setups. There, it is easier to
avoid detrimental averaging over extended sets of geometries,
and additional control parameters such as the orientation of the
dipole moments with respect to the axis of a one-dimensional
sample allow to study the system properties in more detail.



\begin{thebibliography}{99}

\bibitem{book-agarwal}
G.~S. Agarwal,  in {\em Quantum Statistical Theories of Spontaneous Emission
  and Their Relation to Other Approaches}, edited by G. H\"ohler (Springer,
  Berlin, 1974).

\bibitem{thiranumachandran}D.~P. Craig and T.~Thirunamachandran, 
{\it Molecular Quantum Electrodynamics}, Academic, New York, 1984.

\bibitem{book-ficek}Z. Ficek and S. Swain, {\it Quantum Interference and Coherence: Theory
and Experiments}, Springer, Berlin, 2004.

\bibitem{FiTa2002}
Z. Ficek and R. Tanas, Phys. Rep. {\bf 372},  369  (2002).


\bibitem{jump}
M. Lewenstein and J. Javanainen, Phys. Rev. Lett. {\bf 59},  1289  (1987);
A. Beige and G.~C. Hegerfeldt, Phys. Rev. A {\bf 59},  2385  (1999).

\bibitem{Fi1991}
Z. Ficek, Phys. Rev. A {\bf 44},  7759  (1991).

\bibitem{VaAg1992}
G.~V. Varada and G.~S. Agarwal, Phys. Rev. A {\bf 45},  6721  (1992).

\bibitem{Ja1993}
D.~F.~V. James, Phys. Rev. A {\bf 47},  1336  (1993).

\bibitem{RuFiDa1995}
T.~G. Rudolph, Z. Ficek, and B.~J. Dalton, Phys. Rev. A {\bf 52},  636  (1995).

\bibitem{bargatin}
I.~V. Bargatin, B.~A. Grishanin, and V.~N. Zadkov,
Phys. Rev. A {\bf 61}, 052305 (2000).

\bibitem{entanglement}
M.~D. Lukin and P.~R. Hemmer, Phys. Rev. Lett. {\bf 84},  2818  (2000);
G.-x. Li, K. Allaart, and D. Lenstra, Phys. Rev. A {\bf 69},  055802
 (2004).


\bibitem{MaKe2003}
M. Macovei and C.~H. Keitel, Phys. Rev. Lett. {\bf 91},  123601  (2003).

\bibitem{chang}J.-T. Chang, J. Evers, M. O. Scully and M. S. Zubairy,
Phys. Rev. A {\bf 73}, 031803(R) (2006); J.-T. Chang, 
J. Evers and M. S. Zubairy, {\it ibid.} {\bf 74}, 043820 (2006).

\bibitem{AgPa2001}
G.~S. Agarwal and A.~K. Patnaik, Phys. Rev. A {\bf 63},  043805  (2001).

\bibitem{pra-geometry}
J. Evers, M. Kiffner, M. Macovei and C. H. Keitel, 
Phys. Rev. A {\bf 73}, 023804 (2006).

\bibitem{breakdown}
M. Kiffner, J. Evers and C. H. Keitel,
Phys. Rev. A {\bf 76}, 013807 (2007).

\bibitem{dfs}M. Kiffner, J. Evers and C. H. Keitel,
Phys. Rev. A {\bf 75}, 032313 (2007).

\bibitem{strong}M. Macovei, J. Evers, G.-x. Li, and C. H. Keitel,
Phys. Rev. Lett. {\bf 98}, 043602 (2007).

\bibitem{experiment}
R.~G. DeVoe and R.~G. Brewer, Phys. Rev. Lett. {\bf 76},  2049  (1996).

\bibitem{experiment2}
P. Mataloni, E. DeAngelis, and F. DeMartini, Phys. Rev. Lett. {\bf 85}, 
1420 (2000).

\bibitem{experiment3}
J. Eschner, C. Raab, F. Schmidt-Kaler, and R. Blatt, Nature 
{\bf 413}, 495 (2001).

\bibitem{hettich}C. Hettich,  C. Schmitt,  
J. Zitzmann,  S. K\"uhn,  I. Gerhardt, and  V. Sandoghdar,
Science {\bf 298}, 385 (2002).

\bibitem{exp-qdot}P.~Lodahl, A.~Floris van Driel, I.~S.~Nikolaev, 
A.~Irman, K.~Overgaag, D.~Vanmaekelbergh and W.~L.~Vos,
Nature {\bf 430}, 654 (2004).

\bibitem{noel}T.~J. Carroll, K. Claringbould, A. Goodsell, 
M.~J. Lim, and M.~W. Noel, Phys. Rev. Lett. {\bf 93}, 153001 (2004); 
T. J. Carroll, S. Sunder, and M.~W. Noel, 
Phys. Rev. A {\bf 73}, 032725 (2006).

\bibitem{experiment4}
M.~D. Barnes, P.~S. Krstic, P. Kumar, A. Mehta, and J.~C. Wells, 
Phys. Rev. B {\bf 71},  241303(R)  (2005).





\bibitem{time-energy}
M. Kiffner, J. Evers and C. H. Keitel,
Phys. Rev. Lett. {\bf 96}, 100403 (2006).




\end{thebibliography}
\end{document}